\documentstyle[10pt,epsfig]{article}
\newcommand{\myref}[1]{(\ref{#1})}

\newcommand{\mylabel}[1]{\label{#1}}
\newcommand{\mycite}[1]{\cite{#1}}

\begin{document}
\title{New order parameters in the Potts model on a Cayley tree }
\author{F.Wagner, D.Grensing and J.Heide \\
Institut f\"ur Theoretische Physik und Astrophysik \\ der
Christian Albrechts Universit\"at zu Kiel\\
Leibnizstra{\ss}e 15, 24098 Kiel, Germany}
\date{\today}
\maketitle

\begin{abstract} \label{sec_a}
For the $q-$state Potts model
new order parameters projecting on a group of spins instead
of a single spin are introduced. On a Cayley tree  this allows the physical
interpretation of the Potts model at noninteger values $q_0$  of the number
of states. The model can be solved recursively. This recursion
exhibits chaotic behaviour changing qualitatively at critical values
of $q_0$ . Using an additional order parameter belonging to a group of 
zero extrapolated size the additional 
ordering is related to a percolation problem.  This percolation  distinguishes
different phases and explains the critical indices of percolation class 
occuring at the Peierls temperature.
\end{abstract}

\section{Introduction} \label{sec_i}
An interesting feature of the $q$-state Potts model \mycite{Wu82,Tsa96}
is the possibility to consider its thermodynamic observables as
analytic functions of the number $q$ of spin states. Extrapolating
$q$ to unphysical values ($q=0$ (\mycite{For72}) or $q=1/2$ (\mycite{Aha78}))
or taking the derivative at $q=1$ (\mycite{Kas69}) the observables of
other models (resistor network resp. dilute spin glass or percolation) 
are obtained. Since this equivalence holds only for
the thermodynamic observables the connection to the dynamics
of the original Potts model is lost. Extrapolation requires 
explicit  expressions of the observables as functions of $q$. 
This can be achieved, if we consider the $q-$state Potts model
on a Cayley tree, which can be solved
with a recursion formalism \mycite{Egg74}-\mycite{WGH20}.
Extrapolation in $q$ can be done in the recursion for the order
parameter to obtain relations to new models.
As an example we mention the result \mycite{WGH20} that 
the percolation case $q=1$ is equivalent to the logistic equation,
which describes the formation of percolation clusters. To interpret  
non-integer values of $q$ we introduce a new order parameter. 
Instead of distinguishing one of the $q$ values of the spin we assume
ordering for a whole group of $q_1$ spins.
A next neighbour interaction favouring equal spins
will also favour spins in the same group. Therefore it is not surprising
that we find the same mean field equations as in the case $q_1=1$,
however with
an effective number of spins $q_0=q/q_1$ and a redefined coupling 
$K(K_0)$ as function of the original next neighbour coupling $K_0$.
Since $q$ and $q_1$ may be chosen large this allows the interpretation
of the $q-$state Potts model
with noninteger $q_0$ in terms of a physical model. Due to
the redefinition of the coupling not all values of $K$ may correspond to
physical values of the original coupling $K_0$. However, an extrapolation
in the coupling is much easier to perform 
than the one in the number of components. An example will be given. At
$q_0$=1 we encounter the logistic equation exhibiting chaotic behaviour
depending on the value of $K$ as shown in \mycite{WGH20}. One of the aims
of the present paper is to study the transition into chaos for real
values of $q_0$.\\
A problem on a Cayley tree occurs if the recursion allows for 
more than one stable solution. On a normal lattice surface effects can
be neglected and the homogeneous system can be characterized by a
free energy per site $f(m)$ as function of the magnetization. The 
fixed point solutions of the recursion
correspond to different minima of $f$. The phase with the
lowest value of $f$ will be stable, the others are thermodynamically
metastable. On a Cayley tree $f$ cannot be obtained without 
further assumptions due to the presence of the border points.
At a fixed point the interior of a large Cayley tree exhibits the same
magnetization for all sites. As first treated in \mycite{Tho82,Bax82}
for $q=2$ and later on generalized for $q \ge 1$ by Peruggi \mycite{Per83} a free
energy per site $f$ can be found by integration. This $f$ describes
the interior of a large Cayley tree and can be taken as representative 
for the infinite Bethe lattice. At zero magnetic field one obtains
\mycite{Per83} from the absolute minimum of $f$ for ferromagnetic
coupling one transition from the disordered state into a magnetized state
and one transition for negative coupling into an antiferromagnetic
state. The transition obtained by the presumably exact Bethe-Peierls
approximation \mycite{Bet35}
corresponds to a spinodal point of a metastable phase.
The picture of a homogeneous free energy seems not to be very natural
on a Cayley tree since most of the sites are either boundary points or 
lie in the transition region. As shown in \mycite{Agu91} and \mycite{WGH20}
the values of the surface magnetizations decide which of the stable fixed
point is adopted. In this picture the previously as metastable discarded
phase appearing at the Bethe-Peierls temperature corresponds to a 
physical phase. To reconcile the recursion on a Cayley tree with the
approach of a homogeneous $f$ various proposals have been made. The trivial
solution \mycite{WGH20} by chosing the boundary conditions such that
$f$ holds also for the  Cayley tree does not look very 
natural. In \mycite{Guj95}
$f$ ought to be computed by the differenc of free energies between the tree
and a slightly modified lattice which is justified only if the solution
is homogeneous. Monroe \mycite{Mon94} conjectured that the most stable fixed
point corresponds to the physical phase. This criterion reproduces at zero
field the phase diagram of Peruggi \mycite{Per83}. The merit of this
criterion is that it applies locally and does not rely on values at the border.
In the present paper we adopt the picture of Aguiar et al.\mycite{Agu91}
and our previous paper \mycite{WGH20}
that any stable fixed point leads to a physical phase. Within this
picture two questions arise. The first question is how different
phases can distinguished by other properties than as their behaviour at
the border. The second question is related to the end points of the
disordered phase. These are second order phase transitions with
critical indices belonging to the percolation class \mycite{Ess71,WGH20}.
To explain this somewhat surprising 
behaviour we introduce besides a normal ordering parameter 
with respect to one value of the spins a second
ordering on an additional group. By extrapolating the size of the
latter to zero we will see that this ordering 
describes a bond percolation problem with
a probalility depending on the first order parameter. At the end
points of the disordered phase also the percolation becomes critical,
which explains the occurence of percolation indices. If there are
different stable fixed points possible, they will lead to different
bond probabilities.  Therefore the corresponding phases 
can be distinguished apart from the boundary
conditions by their different local percolation properties. \\
The paper is organized in the following way. In section \ref{gen} we give
the general formulae, especially for the the new order parameter. The
mean field equation and possible chaotic solutions are discussed
in section \ref{rec}. In section \ref{sign} we present a 
$q-$state Potts model with an
additional long range force allowing a physical interpretation of the
coupling $K$. Section \ref{phases} is devoted to the appearence of
percolation transitions in the $q$-state Potts model
and section \ref{concl} contains our 
conclusions.

\section{General Formalism} \label{gen}
We consider the $q$-state Potts model on a Cayley
 tree with coordination number $z$. 
At each site $x$ exists a spin $\sigma_x$ letting $q$ different values 
(f.e. \begin{math}\sigma_x=1,2,..q\end{math}). The spins interact with their next neighbours
along the bonds $<xy>$ described by the following Hamiltonian
\begin{equation}\mylabel{A1}
-\beta H=K_0 \sum_{<x,y>}\delta_{\sigma_x,\sigma_y} \quad .
\end{equation}
$K_0$ is proportional to the inverse temperature and the interaction strength.
Negative $K_0$ correspond to antiferromagnetic coupling. Partition function and Boltzmann 
distribution of a single spin can be calculated recursively with the methods 
described in \mycite{WGH20}. A Cayley tree can be thought as $z$ branches of 
length $n$ connected at a site with spin $\sigma$.
Since the Hamiltonian \myref{A1} can be written as a sum over the
branches, we consider the partition function 
$T_n(\sigma)$ for a branch where all 
spin summations inside the branch except the first are carried out.
Any ordering appears as a nontrivial $\sigma$ dependence of $T_n(\sigma)$.
For the description of this ordering we decompose the spin values into
$P+1$ groups with size $q_i$ (i=1,2,..,P+1). Each group $i$ is
characterized by a projector $\Delta^i$ with the property
\begin{equation}\mylabel{A3a}
\Delta_{\sigma}^i=
\left\{
\begin{array}{rl} 
1 & \mbox{$\sigma$ in group $i$}  \\
0 & \mbox{otherwise}
\end{array}\right. \quad .
\end{equation}
The group ordering of  $T_n(\sigma)$ can be written as
\begin{equation}\mylabel{A2}
T_n(\sigma)=\sum_{i=1}^{P+1}u_{in} \; \Delta_{\sigma}^i \quad .
\end{equation}
The disordered state corresponds to value of $u_{in}=u_{n}$ independent of $i$.
The conventional ordering with respect to a single value $\bar{\sigma}$ is 
obtained with P=1 and $\Delta^1_{\sigma}=\delta_{\sigma,\bar{\sigma}}$.
This ordering is for the Ising case $q=2$ also the most general one.
Due to the projection property 
$\Delta^i_{\sigma}\Delta^j_{\sigma}=\delta_{i,j}\Delta_{\sigma}$
the recursion formalism for $P=1$
can be used also in the general case \myref{A2}.
The parameters $u_{in}$ obey a coupled system of recursion relations, as shown
in the Appendix. The spin expectation values $\langle \Delta^i \rangle$ lead
to the magnetizations of a site.
For those we use the linear combinations
\begin{equation}\mylabel{A12}
m_i=\frac{1}{q-q_i} \langle q\Delta^i-q_i \rangle
\end{equation}
\begin{equation}\mylabel{A13}
\overline{m}_i=1-\frac{q}{q_i} \langle \Delta^i \rangle \quad .
\end{equation}
The coefficients have been chosen such that $m_i=\overline{m}_i=0$ holds
for a disordered state $\langle \Delta^i \rangle =q_i/q$ and 
$m_i=1$ ( $\overline{m}_i=1$)
for the ordered case where all (none) of the spin values $\sigma$ belong to 
group $i$. 
Among these $2(P+2)$ magnetizations only $P$ are linearly independent.
Expressions of  $\langle \Delta^i \rangle$ in terms of the ratios 
$u_{in}/u_{1n}$ are given in the Appendix.
We are mainly interested in the properties of only  
one ordering ($P=1$) with respect to a group of size $q=q_1$.
As order parameter we use $x_n=1-u_{2n}/u_{1n}$. The general formula 
of the Appendix	\myref{Ap5} reduces to a recursion formula for $x$
\begin{equation}\mylabel{A9}
x_{n+1}=p(x_n) (1-(1-x_n)^{z-1})
\end{equation}
with
\begin{equation}\mylabel{A11}
p(x)=(e^{K_0}-1)/(e^{K_0}-1+q_1+(q-q_1)(1-x)^{z-1}) \quad .
\end{equation}
For $P=2$ we treat only the case where the recursion system decouples.
This happens for the extrapolation $q_3\rightarrow 0$.
In addition to the recursion \myref{A9} we obtain a quadratic recursion
for $y_n=1-u_{3n}/u_{1n}$
\begin{equation}\mylabel{A10}
y_{n+1}=p(x_n)(1-(1-y_n)^{z-1}) \quad .
\end{equation}
For fixed points $x_n=x$ the logistic equation \myref{A10} describes a bond
percolation with propability $p(x)$ \mycite{WGH20}. Even the order parameter
$y$ does no longer appear in $T$ (since $\Delta^3_{\sigma} \rightarrow 0$ 
for $q_3 \rightarrow 0$), the extrapolated value of 
$\overline{m}_3$ remains finite
and can be related to the percolation ordering parameter $m_P$ defined by
\begin{equation}\mylabel{A9a}
m_P=1-(1-y)^{z}
\end{equation}
through the relation 
\begin{equation}\mylabel{A9b}
\overline{m}_3=m_P-(q/q_1-1)m_1(1-m_P)
\end{equation}
with the magnetization belonging to the order parameter $x$ given by
\begin{equation}\mylabel{A9c}
m_1=\frac{q_1(1-(1-x)^{z})}{q_1+(q-q_1)(1-x)^{z}} \quad .
\end{equation}
\myref{A9} will be discussed in detail in the next section.
The relation of percolation to the phase transitions of the $q$-state 
Potts model given by \myref{A10} will be treated in section \ref{phases}.

\begin{figure}[th]
\psfig{figure=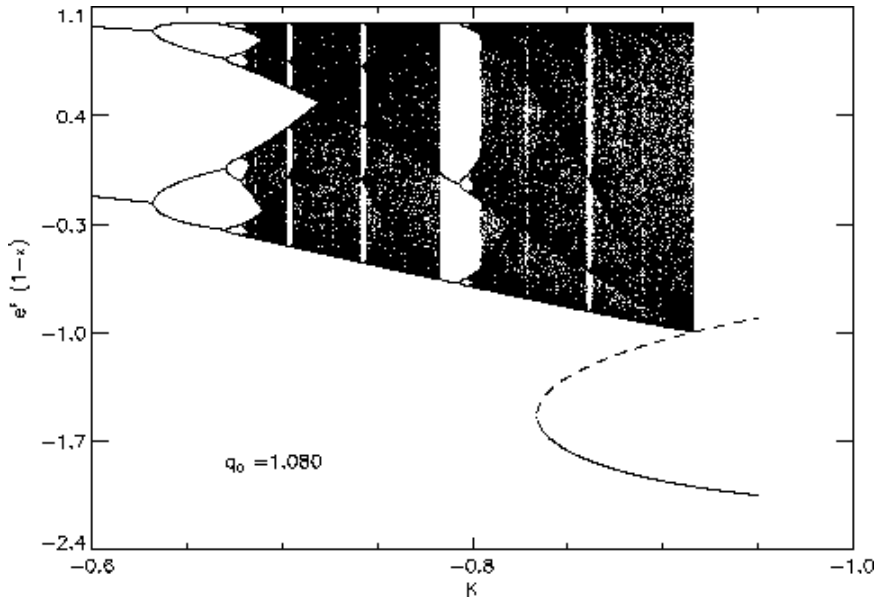,width=11.5cm,angle=0}
\caption{ \label{fig1} Iterates of the recursion \myref{B1} as function of 
          $K$ at $q_0 =1.08$. The curves give the fixed points from
          equ. \myref{B4}.}
\end{figure}

\section{Pottsmodel at non integer values of $q$} \mylabel{rec}
One of the interesting properties of   \myref{A9} is the observation 
that it depends only on two combinations of the three parameters $K_0,q$ 
and $q_1$. If we redefine the inverse temperature $K$ by 
\begin{equation}\mylabel{B1a}
e^K-1=\frac{1}{q_1} (e^{K_0}-1)
\end{equation}
and introduce an effective number of spin states
\begin{equation}\mylabel{B16}
q_0=q/q_1 \quad ,
\end{equation}
\myref{A9} can be written as a mapping
\begin{equation}\mylabel{B1}
x_{n+1}=f(x_n)
\end{equation}
with
\begin{equation}\mylabel{B2}
f(x)=\frac{(e^K-1)(1-(1-x)^{z-1})}{e^K+( q_0  -1)(1-x)^{z-1}} \quad .
\end{equation}
The recursion \myref{B1} is identical to a problem with 
a standard ordering parameter
$q_1=1$ at temperature $K$ with the difference that the 
number of states $q_0$ can 
approximate any real number $q_0\ge1$ for sufficiently large $q_1\le q$.
Therefore the extrapolation of the $q$-states Potts model to noninteger values
of $q_0$ is equivalent to group ordering. In the range $e^K-1\ge q_0/z$
the only stable orbits of the recursion \myref{B1} are fixed points. 
These correspond to 
various phases to be discussed in section \ref{phases}. 
In the remainder of this section we discuss \myref{B1} for real 
negative $K$ and odd $z$. For simplicity we take $z=3$.\\
For $q_0=1$ the mapping \myref{B1} corresponds to the logistic equation,
which exhibits deterministic chaos for the control parameter $e^K$ for the
range $1/2 \le e^K \le 2/3$. For small deviations of $q_0$ from $1$ 
we expect the mapping \myref{B1} to have a similar behaviour. In 
figure  \myref{fig1}  we show the iterates of \myref{B1} for large $n$
as function of $K$ at $q_0=1.08$. This bifurcation diagram is similar to that
of the logistic equation \mycite{Sch84}. In the chaotic regions 
any $x_n$ is possible
within the bounds
\begin{equation}\mylabel{B3}
 f(1-e^{-K}) \le (1-x) \le e^{-K} \quad .
\end{equation}
However, there exists
a characteristic difference to the logistic equation.
For the latter all iterates tend to $\infty$ for $e^{-K} \ge 2$, whereas 
for \myref{B1} the
chaotic regime ends. This can be explained by the existence of a pair 
of fixed points of \myref{B1}:
\begin{equation}\mylabel{B4}
x_{\pm}=1+\frac{1}{2 ( q_0 -1)} (1-e^K \pm \sqrt{(1-e^K)^2-4( q_0 -1)}) 
\end{equation}
possible in the range
\begin{equation}\mylabel{B3a}
|e^K -1| \ge 2 \sqrt{ q_0  -1} \quad .
\end{equation}  
$x_+$ is stable (solid line in fig \myref{fig1}) and $x_-$ 
(broken line) is unstable for $K \le 0$.
The bassin of attraction of $x_+$ is given by initial values $x_{0} \ge x_-$.
Therefore the chaotic regime ends where the lower bound in \myref{B3} is
equal to $x_-$ from \myref{B4}. Solving this equation for $K$ we find
the endpoint $K_E$ of the chaotic regime
\begin{equation}\mylabel{B5}
e^{K_E}=\frac{1}{4} (1+\sqrt{9-8 q_0  }) \quad .
\end{equation}

\begin{figure}[ht]
\psfig{figure=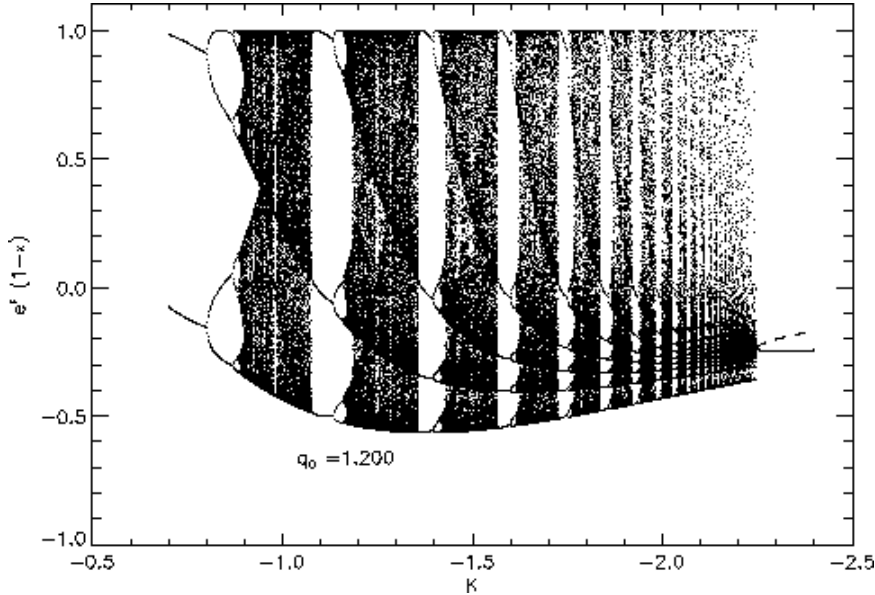,width=11.5cm,angle=0}
\caption{ \label{fig2} Iterates of the recursion \myref{B1} as function of 
          $K$ at $q_0 =1.20$. Chaotic regions are intercepted by stable
          orbits with increasing periods. The accumulation point of these
          orbits is given by equ \myref{B3a}.}
\end{figure}
If we increase $q$ beyond $10/9$ the endpoint of the chaotic regime is 
determined by equality in \myref{B3a}.
At the same time the pattern of the chaotic behaviour changes qualitatively.
An example for $ q_0 =1.2$ is shown in figure \myref{fig2}. 
The regions of chaos are intercepted by stable orbits with 
increasing periods of $\tau=2,3,..$. If we enlarge
the region in between the stable orbits they are similar to the pattern
seen in figure \myref{fig1}. The stable orbits exhibit an accumulation point 
at the end of the chaotic region given by equality in \myref{B3a}.
The period $\tau$ of the orbit as function of $K$ may be estimated 
(for sufficiently large $\tau$) by replacing the recursion formula by a
differential equation:
\begin{figure}[ht]
\psfig{figure=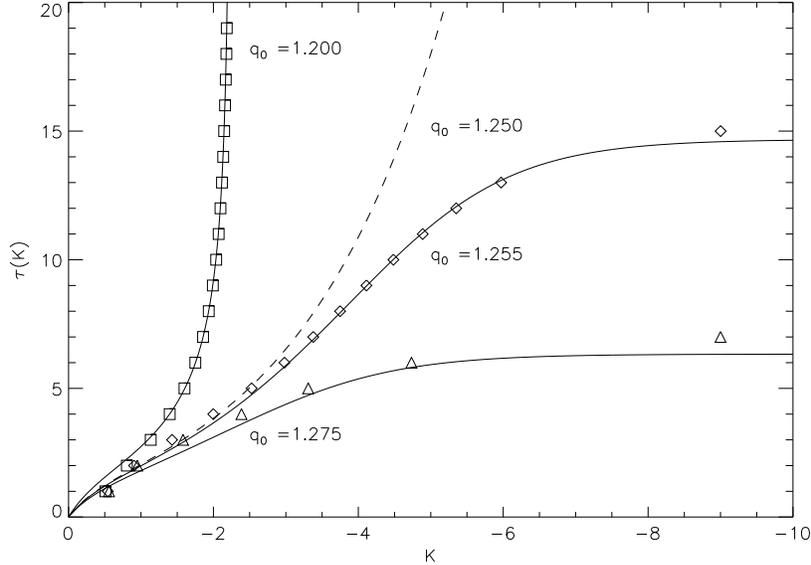,width=11.5cm,angle=0}
\caption{ \label{fig3} The continuum period \myref{B7} as function of 
          $K$ for various values of $q_0$. The data points are
          estimates from figures \myref{fig2},\myref{fig4} and \myref{fig5}
          using as $K$ value the locus of the first period doubling.
          The dashed line gives the critical curve for $q_0 =5/4$. }
\end{figure}
\begin{equation}\mylabel{B6}
\frac{dx}{dn} \approx x_{n+1}-x_n=f(x)-x \quad .
\end{equation}
The solution $x(n)$ of \myref{B6} can be found by an elementary integration.
They are periodic with a period
\begin{equation}\mylabel{B7}
\tau=\frac{2 \pi (1-e^K)(2-q_0)}{(1-e^K+q_0)+\sqrt{4(q_0-1)-(e^K-1)^2}} \quad .
\end{equation}
In figure \myref{fig3} we compare this $\tau(K)$ with the integer length 
of the orbits in figure \myref{fig2}. As corresponding $K$ value we take 
the first bifurcation in each intervall.
As one sees from figure \myref{fig3}
 the agreement between the observed periods and the estimate 
\myref{B7} is excellent, already at small $\tau$. 
For values of $q_0$  larger than $5/4$ the fixed points \myref{B4}
 no longer exist. The periods as seen in figures \myref{fig4} ($q_0 =1.255$)
and \myref{fig5} ($q_0 =1.275$) remain finite. This is confirmed by the 
agreement with \myref{B7} in figure \myref{fig3}. The last chaotic region
extends to $K\rightarrow-\infty$ and the number of regions of stable orbits 
decrease with increasing $q_0$ . For $ q_0  > q_c$ with $q_c \approx 1.53$
there will be no transition into chaos. For $1.56 \le  q_0  \le 3$ 
only a $\tau =2$ orbit exists. For $ q_0  \ge 3$ and negative coupling $K$ we 
have only the disordered fixed point $x=0$.\\
\begin{figure}[ht]
\psfig{figure=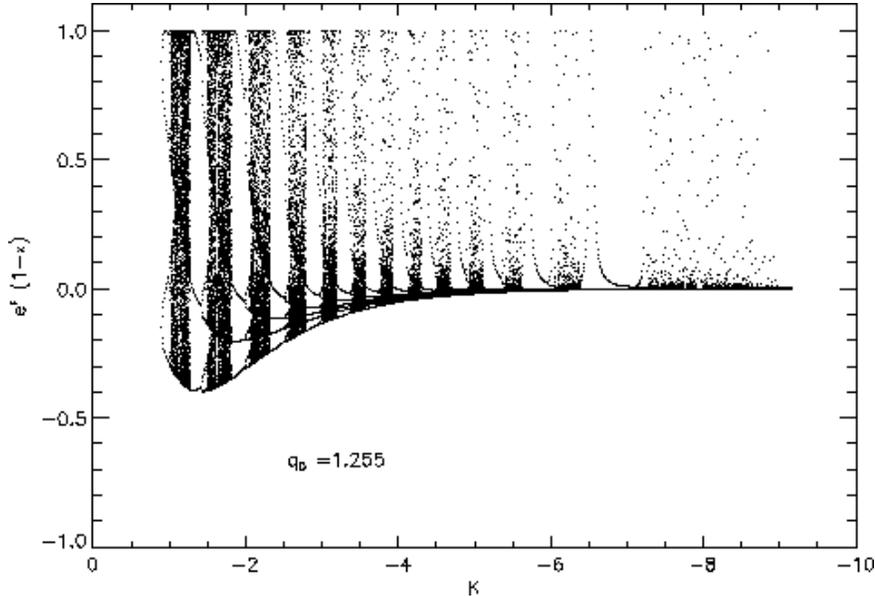,width=11.5cm,angle=0}
\caption{ \label{fig4} Iterates of the recursion \myref{B1} as function of
          $K$ at $q_0 =1.255$. Different chaotic regimes are intercepted
          by a finite number of stable orbits.}
\end{figure}
We see that both the recursion \myref{B1} for general 
$q_0$ and the logistic
equation for $q_0 =1$ have a control parameter $K$ 
which decides on the details of
the chaotic behaviour. In addition the parameter $q_0 =q/q_1$ decides which 
type of chaotic behaviour occurs. The various types we found are summarized
in table \myref{table}.
\newpage
\begin{figure}[th]
\psfig{figure=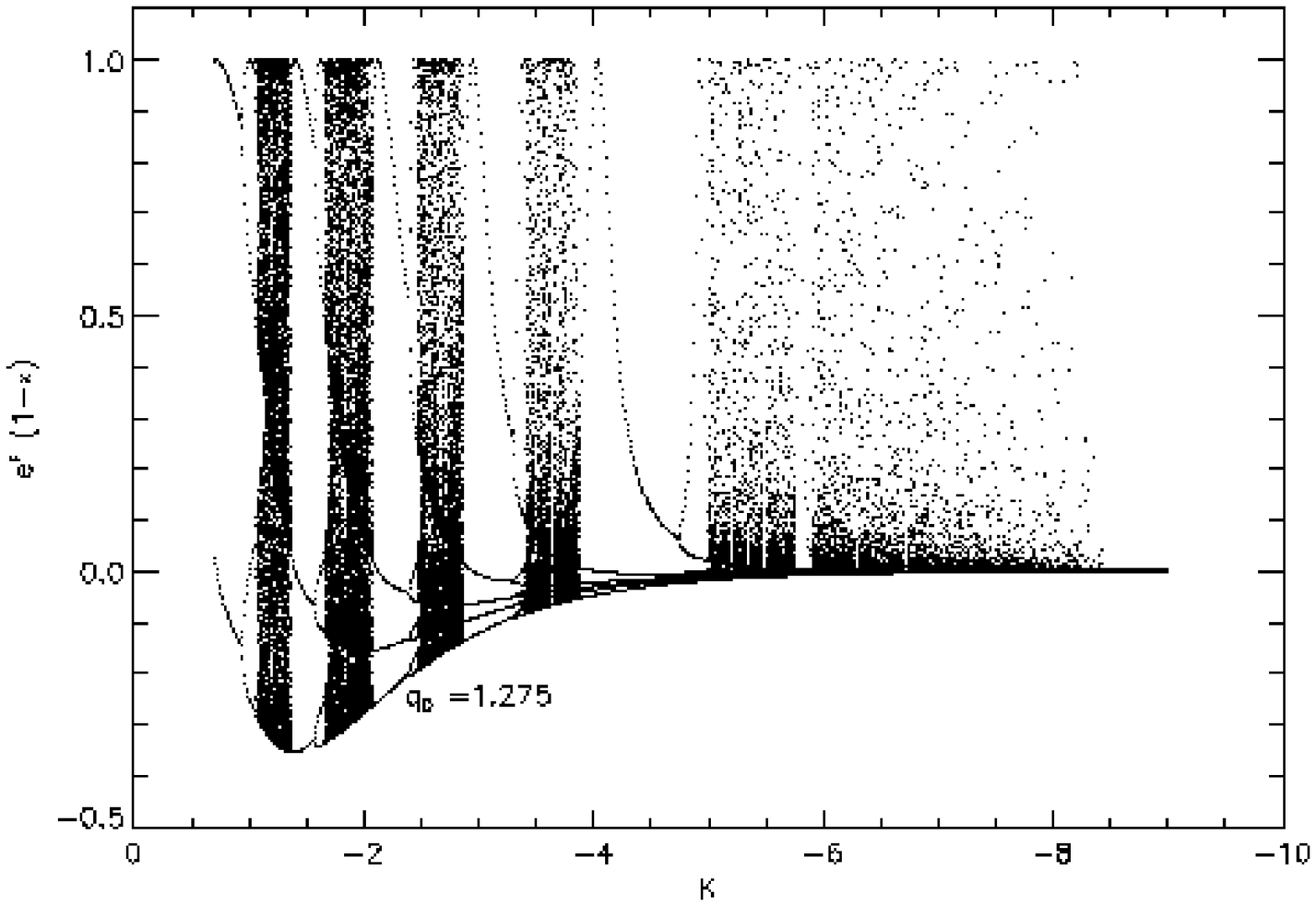,width=11.5cm,angle=0}
\caption{ \label{fig5} Iterates of the recursion \myref{B1} as function of
          $K$ at $q_0 =1.275$. Different chaotic regimes are intercepted
          by a finite number of stable orbits.}
\end{figure}
\begin{table}[hb] 
\begin{tabular}{|c|c|c|c|} \hline
q range & end point $\exp(K)$ of & repetition of & maximum\\
        & chaotic behaviour & deterministic chaos & orbit length\\ \hline
\rule[ 1.5ex]{0mm}{1.5ex} 
$1 \le  q_0  \le 10/9 $& $1/4(1+\sqrt{9-8  q_0 }) $ & 1 & $\infty$ \\
\rule[ 1.5ex]{0mm}{1.5ex} 
$10/9 <   q_0  \le 5/4$ & $1-2 \sqrt{  q_0 -1}$ & $\infty$ & $\infty$\\
\rule[ 1.5ex]{0mm}{1.5ex} 
$5/4<  q_0 <q_c$ & $0$ & $2 \pi (2-  q_0 )/(  q_0 +1)\sqrt{4  q_0 -5}$ 
                          & $\infty$\\
\rule[ 1.5ex]{0mm}{1.5ex} 
$  q_0 >q_c$ & - & $0$ & finite \\
\rule[ 1.5ex]{0mm}{1.5ex} \rule[-1.5ex]{0mm}{1.5ex} 
$1.56<  q_0 <3$ & - & 0 & 2 \\ \hline
\end{tabular}
\caption{\label{table} Characteristics of chaotic behaviour in different
         intervalls of $q_0$. The value of $q_c$ above which no orbit
         of infinite length occurs is given numerically by $q_c=1.53$.}
\end{table}

\section{A Generalized Potts model} \mylabel{sign}
As shown in the previous section a Potts model 
extrapolated to non integer values 
$q_0$ is equivalent to a $q$-state Potts model at a rescaled temperature using
an order parameter describing simultaneous ordering of a group of $q_1$
spins. The restriction to physical values of $K_0$ imposes by \myref{B1a}
a lower limit of $K$, which excludes the chaotic behaviour discussed in section
\ref{rec}. Therefore still an extrapolation is required. 
To find a  model, where $K$ has a 
physical meaning, we introduce the number $E$ of occupied links in each
spin configuration: 
\begin{equation}\mylabel{c1}
E= \sum_{<x,y>}\delta_{\sigma_x,\sigma_y} \quad .
\end{equation}
A new model may be defined by allowing only spin
configurations where $ E      $ is even
or odd. This corresponds to a nonlocal spin interaction. The partitioned sum
for even $E$ then will involve branch contributions with both even
and odd $E$. It is convenient to define partitioned sums $T_n^{e}(\sigma)$ or
$T_n^{o}(\sigma)$ for branches allowing only even or odd values for $E$
in the branch. As in the previous case we express $T_n^{e,o}(\sigma)$
by contributions of the subbranches $T_{n-1}^{e,o}(\sigma)$. This system
decouples, if we consider the sum and difference of even and odd branches
given by
\begin{equation}\mylabel{C2}
T_n^{(\pm )}(\sigma)=T_n^{e}(\sigma)\; \pm \; T_n^{o}(\sigma) \quad .
\end{equation}
If we parametrize $T_n^{(\pm )}$ in the form \myref{Ap2}
\begin{equation}\mylabel{C3}
T_n^{(\pm )}(\sigma )=\sum_{i}^{P+1}u_{i,n}^{(\pm )}\Delta^i_{\sigma} \quad ,
\end{equation}
$T_n^{(+)}$ corresponds to the unrestricted case given by \myref{Ap2}
and \myref{Ap4} in the Appendix.
Using only one order parameter ($P=1$) and introducing the ratio
\begin{equation}\mylabel{C4}
x_n^{(-)}=\frac{\displaystyle u_{2,n}^{(-)}}{u_{1,n}^{(-)}}\; -\; 1 \quad ,
\end{equation}
we obtain the recursion formula
\begin{equation}\mylabel{C5}
x_{n+1}^{(-)}= \frac{\displaystyle (1+e^{K_0})((1-x_n^{(-)})^{z-1}-1)}
                    {{q}_1 -1-e^{K_0}+(q-{q}_1)(1-x_n^{(-)})^{z-1}}  \quad .
\end{equation}
Redefining the temperature by
\begin{equation}\mylabel{C6}
e^{K_0}+1={q_1}(1-e^K)  \quad ,
\end{equation}
we obtain with $q_{0}=q/q_1$ the same recursion formula as \myref{B1} 
without restriction on $K$. The chaotic region of \myref{B1} can be
reached for large $q_1$ and fixed value for $K_0$. In this
range the according  recursion to $x^{(+)}_n$ leads always to the
trivial stable fixed point $x^{(+)}=0$.

\section{Phases of the Potts model } \label{phases}
The stable fixed points of recursion \myref{A9} give 
in our interpretation the possible
phases with respect to the group ordering parameter $x$. We do
not perform the replacements \myref{B1a} and \myref{B16}, but
describe the model in terms of the original parameters $K_0,q$ and
$q_1$. There exist three critical points $K_c,K_c^\prime$ and
$K_c''$. The disordered fixed point $x=0$ is stable  at
high temperatures in the 
range $K_c'  <  K_0  <  K_c$ with $K_c'=\ln (1-q/z)$ and
$K_c=\ln (1+q/(z-2))$. As the temperature is lowered two new  fixed points
given by \myref{B4} appear for $K_0  >  K_c''$. Only $x_+$ with
positive magnetization $m_1$ corresponds to a stable fixed point.
For $z=3$ the value of $x_+ $ is given by equ. \myref{B4}. In contrast to 
$K_0$ and $K_0'$ the value of $K_c''$ depends on $q_1$. For $z=3$
one finds $K_c''=\ln (1+2\sqrt{q_1 (q-q_1 )})$.
Therefore there exists a series of critical points depending on
$q_1$. For $ K_0  >  K_c$ the disordered phase changes into the
fixed point $x_-$ ( for $z=3$ given by equ. \myref{B4} ) with
negative magnetization $m_1$, and for $ K_0  <  K_c'$ we encounter
an antiferromagnetic coupling. This transition corresponds
to the first bifurcation discussed in section \ref{rec}. If
more than one stable fixed point exists, the value at the boundary
decide about the adopted phase. The transitions at $K_c''$ is of
first order. For the following it does not matter wether this
transition apears at $K_c''$ or a higher value as advocated in
\mycite{Per83} from equality of the free energies. It is only important
that the transitions at $K_c$ and $K_c'$ are genuine transitions
and do not correspond to spinodal points.
For the critical indices at $K_c$ we
find values belonging to the percolation class ($\beta =\gamma =1$).
The same is true for the average magnetization also at $K_c'$.	
The staggered magnetization equal to the
difference of magnetizations at adjacent sites exhibits
at $K_c'$ indices of the mean field class ($2 \beta =\gamma =1$).
The Ising case $q_1 =q/2$ is exceptional. Since the amplitude
of the average magnetization vanishes due to the Ising symmetry
on an $AB$-lattice ($m(x_+)+m(x_-)=0$) the percolation transitions
dissappear, and one observes mean field indices at both critical
points $K_c'$ and $K_c''=K_c$. Note that a Ising type behaviour
does not only occur for $q=2q_1=2$ but also for any even $q$
with the choice $q_1=q/2$. This reflects the fact that the
recursion \myref{A9} can be written in the form \myref{B1}
which depends only on the ratio $q_0=q/q_1$.\\
The appearance of percolation indices for $q_1\not=q/2$ is
somewhat surprising. To understand the mechanism we consider the
recursion \myref{A9} for the order parameter $x$ together with an
extrapolated second order parameter $y_n$ given by \myref{A10}.
The recursion for the latter  reads at a fixed point for $x$ as
\begin{equation} \mylabel{D1}
y_{n+1}=p(x)(1-(1-y_n)^{z-1})
\end{equation}
with $p(x)$ given by \myref{A11}. As shown in \mycite{WGH20} the recursion
\myref{D1} is equivalent to bond percolation with a probability $|p(x)|$.
A non zero fixed point value  $y$ leads to a finite probability $m_P$
for a site belonging to the infinite cluster
\begin{equation} \mylabel{D4}
m_P=1-(1-y)^z \quad .
\end{equation}
The mapping \myref{D1} has the 
fixed points $y=0$ and $y=x$. The stable fixed point must satisfy
\begin{equation} \mylabel{D3}
|p(x)(z-1)(1-y)^{z-2}|<1 \quad .
\end{equation}
Both the disordered fixed point $x=0$ in the range $K_c'<K_0<K_c$
 and $x_-$ with negative magnetization for $K_0>K_c$ lead to the 
nonpercolating state $y=0$. In contrast the fixed point $x_+$
requires $y=x_+$. Therefore the two phases possible for $K_0>K_c''$
can be distinguished locally by their percolating properties.
The phase with $m_1>0$ exhibits an infinite percolation cluster,
whereas the phase with $m_1 \le 0$  is in the disordered percolation state.
Using the criterion of existence of an infinite cluster the phase
on the Cayley tree is uniquely defined and can be carried over to the infinite
Bethe lattice, since  one does no  longer need the values at the boundary.
For $K_0<K_c'$ the recursion \myref{A9} exhibits an $\tau=2$ orbit
corresponding to antiferromagnetic ordering. By a straightforward
calculation one can show that the recursion \myref{A10} has a stable
fixed point $y=0$, although $p$ alternates between $p(x_{\pm})$ in
subsequent iterations.\\
If $K_0$ approaches the end points $K_c$ or $K_c'$ of the disordered phase,
$|p(x)|$ approaches the critical value $1/(z-1)$. At these points
all moments of the cluster size distribution diverge. These moments
can be represented by
derivatives of the magnetization $m_P$ with respect to an external field.
Therefore magnetization and susceptibility will be singular
 with critical indices belonging to the 
percolation class. Since the various magnetizations are not independent
due to relation \myref{A9b} 
these divergences will show up also in other magnetizations than $m_P$.

\section{Conclusions} \label{concl}
We introduced a new type of ordering in the $q-$state Potts model
where a spin belongs to
one or more groups of values instead of the conventional ordering with
respect to one value. 
The recursion formalism for the latter can be also applied for the new ordering.
For only one group of size $q_1$ the recursions
 can be cast into the same form as
for $q_1=1$ with the exception that the effective number of states is 
given by $q_0=q/q_1$, which can approximate any real number larger than 1.
This allows an extrapolation of the $q-$state Potts model
to noninteger values of $q$
without losing the physical interpretation. For $q_0$ close to 1 the
recursion exhibits deterministic chaos similar to the logistic
equation. The control
parameter is essentially the temperature $K^{-1}$. Increasing $q_0$
 we observe as function of $K$ repetitions of deterministic chaos separated
by stable orbits of length $\tau=2,3,4,\cdots$. Number of the orbits and their
length can be estimated by a continous length approximation. Most of this
chaotic behaviour occurs at $K(K_0)$ values, which cannot be reached with
real values of $K_0$. To avoid this extrapolation to unphysical values
of the temperature one can introduce a generalized Potts model, where
possible spin configurations  are restricted 
by the condition that the number of links 
connecting sites with equal spin values must be even or odd.
In \mycite{Per87} the Potts model has been extrapolated to
real values of $q$ using the analytic expression for the free energy.
For $q<2$ negative values for the specific heat have been found. This
unphysical behaviour can be avoided if one uses the new order parameter
referring to a group of $q_1$ spins and interprets the ratio $q_0=q/q_1$
as effective number of spins.\\
 The phase structure of the $q-$state Potts model
with ordering with respect to one
group of size $q_1$ are qualitatively similar to the conventional
$q_1=1$ case. Only for even $q$ the Ising case $q_1=q/2$ behaves
differently. For $q_1 \neq q/2$ the transitions at the endpoints of the
disordered phase are of second order with percolation class critical 
indices. To understand their occurence we added a second group of
ordering with size extrapolated to zero. This does not change the
recursions for the first group. In addition one obtains a logistic 
equation for the second
order parameter, which leads to indices of the percolation class. Since
the control parameter in the logistic equation
 depends on the phase, different phases can be
distinguished by their percolation property. The phases with positive
magnetization exhibit an infinite cluster with finite probability, whereas
negative or zero magnetizations $\leq 0$ have zero probability. 
This criterion allows a local distinction
between two possible phases. Since one does not have to resort to
boundary conditions, this criterion can be applied also on the infinite 
Bethe lattice.\\
 
Acknowledgements: One of uf us (J.H) likes to thank the Landesbank of 
Schleswig-Holstein for financial support.

\newpage
\listoffigures

\begin{appendix}
\section{Recursion on a Cayley tree}
The partitioned sum for a Cayley tree can decomposed 
into $z$ branches $T_n(\sigma )$ of length $n$ connected at a central site
with spin $\sigma $. Each branch can be related according 
to figure \myref{tree}  to $z-1$ branches  of length $n-1$ by
\begin{equation}\mylabel{Ap1}
T_n(\sigma )=\sum_{\sigma '}\exp (K_0\delta _{\sigma ,\sigma '})
             \left( T_{n-1}(\sigma ')\right)^{z-1}
\end{equation}
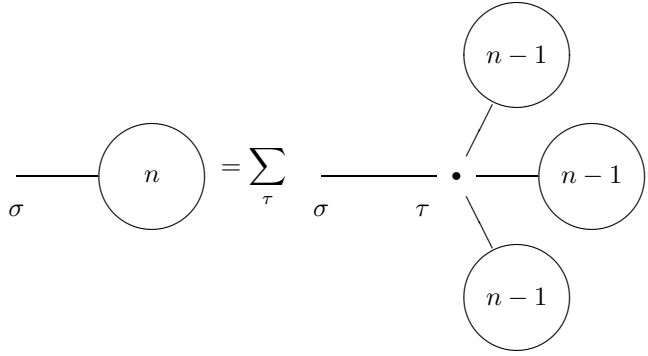
\begin{figure}[h]
\setlength{\unitlength}{0.9cm}
\begin{picture}(9,5.5)
\put( 0.5,2.5){\makebox(0.0,0.0){$\sigma $}}
\put( 0.5,3.0){\line(1,0){1.2}}
\put( 2.5,3.0){\circle{2.0}} \put( 2.5,3.0){\makebox(0.,0.){$n$}}
\put( 4.2,3.0){\makebox(0.0,0.0){$\displaystyle =\sum_{\tau} \quad$}}
\put( 5.0,2.5){\makebox(0.0,0.0){$\sigma $}}
\put( 5.0,3.0){\line(1,0){1.7}}
\put( 7.0,3.0){\circle*{0.1}}  \put( 6.5,2.5){\makebox(0.,0.){$\tau$}}
\put( 7.3,3.0){\line(1,0){0.9}}
\put( 9.0,3.0){\circle{3.5}} \put( 9.0,3.0){\makebox(0.,0.){$n-1$}}
\put( 7.15,3.3){\line(1,2){0.38}}
\put( 7.89,4.79){\circle{3.5}} \put( 7.89,4.79){\makebox(0.,0.){$n-1$}}
\put( 7.15,2.7){\line(1,-2){0.38}}
\put( 7.89,1.21){\circle{2.5}} \put( 7.89,1.21){\makebox(0.,0.){$n-1$}}
\end{picture}
\caption{ \label{tree} Recursion for $z=4$}
\end{figure}
Parametrizing $T_n(\sigma )$ with
\begin{equation}\mylabel{Ap2}
T_n(\sigma )=\sum_{i}u_{i,n}\Delta^i_{\sigma}
\end{equation}
the products in \myref{Ap1} can be evaluated using the relation
\begin{equation}\mylabel{Ap3}
\Delta^i_{\sigma}\Delta^j_{\sigma}=\delta_{ij}\Delta^i_{\sigma}
\end{equation}
Comparing coefficients on both sides we find
\begin{equation}\mylabel{Ap4}
u_{i,n}=\left(e^{K_0}-1\right)\left(u_{i,n-1}\right)^{z-1}\; +\;
        \sum_{j}q_j\left(u_{j,n-1}\right)^{z-1}
\end{equation}
Spin expectation values can depend only on the ratios of $u_{i,n}$.
For these we obtain the recursion
\begin{equation}\mylabel{Ap5}
\frac{\displaystyle u_{i,n}}{u_{1,n}}=1\;+\; 
      \frac{\displaystyle \left ( e^{K_0}-1 \right )
      \left ( \left ( u_{i,n-1}/u_{1,n-1}\right)^{z-1} -1 \right)}
      {e^{K_0}-1  + \sum_{j}q_j\left(u_{j,n-1}/u_{1,n-1} \right)^{z-1}}
\end{equation}
Setting $u_{2,n}/u_{1,n}=1-x_n$ and $u_{3,n}/u_{1,n}=1-y_n$ we obtain 
at $q_3=0$ the recursions \myref{A9} and \myref{A10} in the case P=2.
The Boltzmann weight for the central spin at a fixed point $r_i=u_i/u_1$
is given by 
\begin{equation}\mylabel{Ap6}
w(\sigma )=N\left(T(\sigma ) \right)^{z}
\end{equation}
where the normalization constant follows from $\sum w(\sigma )=1$. With
the distribution \myref{Ap6} we get for $\langle \Delta^i \rangle$ 
\begin{equation}\mylabel{Ap7}
\langle \Delta^i \rangle=\frac{\displaystyle q_i(r_i)^z}
                              {q_1+\sum_{j>1}q_j(r_j)^z}
\end{equation}
If we use in the case $P=2$ as independent
magnetizations $m_1$ and $\overline{m}_3$, we find
\begin{equation}\mylabel{Ap8}
m_1=q_1\left(1-(1-x)^z\right) /N
\end{equation}
and
\begin{equation}\mylabel{Ap88}
\overline{m}_3=1-q(1-y)^z/N
\end{equation}
with the normalization factor $N=q_1+(q-q_1)(1-x)^z$.
\end{appendix}

\end{document}